\documentclass[twocolumn,prc,aps,showpacs]{revtex4}
\usepackage{graphicx}

 \def\Pom{{ I\!\!P}}
 
 \def\gsim{\mathrel{\rlap{\lower4pt\hbox{\hskip1pt$\sim$}}
 \raise1pt\hbox{$>$}}}

 \newcommand\la{\langle}
 \newcommand\ra{\rangle}
 \newcommand\beq{\begin{equation}}
 
 \newcommand\eeq{\end{equation}}
 \newcommand\beqn{\begin{eqnarray}}
 \newcommand\eeqn{\end{eqnarray}}
 \newcommand\nn{\nonumber}
\def\mb{\,\mbox{mb}}

\def\GeV{\,\mbox{GeV}}

\def\nto{\to\!\!\!\!\!\!/\ \ }
\def\lsim{\mathrel{\rlap{\lower4pt\hbox{\hskip1pt$\sim$}}
    \raise1pt\hbox{$<$}}}         
\def\gsim{\mathrel{\rlap{\lower4pt\hbox{\hskip1pt$\sim$}}
    \raise1pt\hbox{$>$}}}         

\def\Im{\,\mbox{Im}\,}
\def\mb{\,\mbox{mb}}

\def\GeV{\,\mbox{GeV}}

\begin{document}

\title{\bf Breakdown of PCAC in diffractive neutrino interactions}

\vspace{1cm}

\author{B. Z. Kopeliovich}
\author{I. K. Potashnikova}
\author{Iv\'an Schmidt}
\author{M.~Siddikov}
\affiliation{Departamento de F\'{\i}sica,
Universidad T\'ecnica Federico Santa Mar\'{\i}a; and
\\
Instituto de Estudios Avanzados en Ciencias e Ingenier\'{\i}a; and\\
Centro Cient\'ifico-Tecnol\'ogico de Valpara\'iso;\\
Casilla 110-V, Valpara\'iso, Chile}
\begin{abstract}
\noindent We test the hypothesis of partially conserved axial
current (PCAC) in high energy diffractive neutrino production of
pions. Since the pion pole contribution to the Adler relation (AR)
is forbidden by conservation of the lepton current, the heavier states,
like the $a_1$ pole, $\rho$-$\pi$ cut, etc., control the lifetime of
the hadronic fluctuations of the neutrino. We evaluate the deviation from 
the AR in diffractive neutrino-production of pions on proton and nuclear targets.
At high energies, when
all the relevant time scales considerably exceed the size of the
target, the AR explicitly breaks down on an absorptive
target, such as a heavy nucleus. In this regime, close to the black disc
limit, the off-diagonal diffractive amplitudes vanish, while
the diagonal one, $\pi\to\pi$, which enters the AR, maximizes
and saturates the unitarity bound. At lower energies, in the regime
of short lifetime of heavy hadronic fluctuations the AR is restored, 
i.e. it is not altered by the nuclear effects.
\end{abstract}



\maketitle

\section{Introduction}

 In the chiral limit of massless quarks isovector components of both
vector and axial quark currents are conserved. Though the hadrons acquire masses
via the mechanism of spontaneous symmetry breaking, hadronic currents are 
still conserved. While it is rather straightforward for the vector current,
conservation of the axial current looks nontrivial, and is
possible only due to the presence of a pseudo-scalar term
in the current, which has to be singular at $Q^2=0$ \cite{nambu}. 
This singularity is associated with massless Goldstone particles \cite{goldstone} or
pions, which appear due to spontaneous chiral symmetry breaking.

Beyond the chiral limit, the pions acquire a small mass and the axial current
conservation is not exact, so one can consider a partial conservation
of the axial current (PCAC),
\beq
\partial_\mu J^A_\mu=m_\pi^2\,f_\pi\,\phi_\pi,
\label{20} \eeq where $m_\pi$ and $f_\pi\approx 0.93 m_\pi$ are the
pion mass and decay coupling, and $\phi_\pi$ is the pion field.

A beautiful manifestation of PCAC is the Goldberger-Treiman relation
\cite{treiman}, which bridges weak and strong interaction. It miraculously
connects the pion decay constant with the
pion-nucleon coupling, which seem to have very little in common.
Indeed, the former depends on the pion wave function, while the latter
is controlled by the wave function of the nucleon.  Nevertheless,
data on $\beta$-decay and muon capture confirm this
relation between very different physical quantities. This astonishing relation between 
the pion pole (suppressed in beta-decay due to conservation of the lepton current)
and heavier states having no natural explanation, except PCAC, was called
Goldberger-Treiman conspiracy \cite{bell-book} (see more below).

Another intensive source of axial current is high energy neutrino
interactions. In this case PCAC leads to the Adler relation (AR)
between the cross sections of processes initiated by neutrinos and
pions \cite{adler},
\beq \left.\frac{d^2\sigma(\nu p\to
l\,X)}{dQ^2\,d\nu}\right|_{Q^2=0} = \xi^2(E,\nu)\,\sigma(\pi p\to
X). \label{40} 
\eeq 
Here the kinematic factor is
 \beq
 \xi^2(E,\nu)=\frac{G^2}{2\pi^2}\,f_\pi^2\,\frac{E-\nu}{E\nu};
 \label{50}
 \eeq
$E$ is the neutrino energy;  $G=1.166\times 10^{-5}\,GeV^{-2}$ is
the electro-weak Fermi coupling; $Q^2=-q_\mu^2$, where
$q_\mu=k_\mu-k^\prime_\mu$ and $\nu=E-E^\prime$ are the 4-momentum
and energy transfer in the $\nu\to l$ transition (the same notation
as for neutrinos should not cause confusion). For the sake of
concreteness the target is the proton, but it may be any hadron or a
nucleus.

A high-energy neutrino exposes hadronic properties interacting via its hadronic fluctuations \cite{goldman}.
Similar to the Goldberger-Treiman relation the AR
(\ref{40}) should not be interpreted as pion pole dominance. 
Neutrino cannot fluctuate to a pion, $\nu\nto\pi
l$, because the pion pole in the dispersion relation in $Q^2$ for
the axial current does not contribute to the interaction of the
neutrino at high energies \cite{bell,bell-book,p-s,km}. 
Indeed, the axial current $J^A_\mu(Q^2)$
can be presented as,
\beqn
J^A_\mu(Q^2)&=&\frac{q_\mu\,f_\pi}{Q^2+m_\pi^2}\,T(\pi p\to X) \nn\\
&+& \frac{f_{a_1}}{Q^2+m_{a_1}^2}\,T_\mu(a_1 p\to X)\,+\,...
\label{100} \eeqn
Here the second and following terms represent the
contributions of the $a_1$ meson and (implicitly) other heavier
axial-vector states.

The first term in (\ref{100}), corresponding to the pion pole,
contains the factor $q_\mu$, which then terminates its contribution
to the cross section, Eq.~(\ref{40}). Indeed, the amplitude of the
reaction is
\beq A(\nu\,p\to l\,X) \propto L_\mu\,J^A_\mu,
\label{120} \eeq
where $L_\mu=\bar
l(k^\prime)\gamma_\mu(1+\gamma_5)\nu(k)$ is the lepton current,
which is transverse, i.e.  $q_\mu\,L_\mu=0$
(for simplicity hereafter we entirely neglect the lepton mass).
Therefore,
the pion term in (\ref{100}) does not contribute to the amplitude
Eq.~(\ref{120}), and this is true at any $Q^2$.

Thus, although it is tempting to interpret the AR Eq.~(\ref{40}) as
a manifestation of the pion pole dominance, this is not correct. 
PCAC connects the contribution of heavy axial
states  (the second line in Eq.~(\ref{100})) with the nonexistent
pion contribution at $Q^2=0$ 
\cite{bell,bell-book,p-s,km}. Such a fine tuning, which is very similar to 
the Goldberger-Treiman conspiracy,  looks miraculous,
and the PCAC hypothesis for neutrino interactions should be tested
thoroughly.

A simple way to see in data 
whether light or heavy states dominate the dispersion
relation for the axial current, is to measure the $Q^2$-dependence 
of the neutrino cross section at small $Q^2$.
Extrapolating the cross section Eq.~(\ref{40}) with the parametrization 
$(Q^2+M_{eff}^2)^{-2}$, one can find the position
$Q^2=-M_{eff}^2$ of the essential singularity in the dispersion
relation (\ref{100}). It is easy to disentangle between the
effective masses which are as small as the pion mass and heavy
singularities, like the $\rho$-$\pi$ cut, $a_1$ meson, etc. Data
clearly prefer the latter, $M_{eff}\gsim 1\GeV$ \cite{km}.

\section{Diffractive neutrino-production of pions}\label{sect2}

This reaction offers probably the most stringent test of PCAC in
neutrino interactions. Indeed, the analysis performed by Piketty and
Stodolsky \cite{p-s} revealed a potential problem related to the
above dispersion representation for the AR. They made use of the 
relation between the pion pole and heavy states contribution in Eq.~(\ref{100}) 
imposed by PCAC, complemented with few assumptions. 
The assumed dominance of the axial vector $a_1$ meson, i.e. neglected
the higher terms implicitly contributing in (\ref{100}).
Also assumed a smooth $Q^2$-dependence of the hadronic amplitudes
in (\ref{100}), and related the lepton coupling $f_{a_1}$ to that for $\rho$ meson  
relying on the Weinberg sum rules \cite{weinberg}. Eventually,  they
arrived at a relation between the elastic and diffractive
pion-nucleon cross section, $\sigma(\pi p\to a_1 p)\approx\sigma(\pi
p\to\pi p)$. This relation strongly contradicts data: diffractive
production of $a_1$ meson is more than an order of magnitude
suppressed compared with the elastic cross section.

This puzzle was relaxed in \cite{belkov,km} by pointing out its shaky point,
namely, the $a_1$ pole cannot dominate in the axial current, since
it is quite a weak singularity compared to the $\rho$ pole in the
vector current. In fact, the main contribution to the expansion
Eq.~(\ref{100}) comes from the $\rho$-$\pi$ cut, related
to diffractive pion excitations. The invariant mass distribution
for diffractive $\pi\to3\pi$ excitations peaks at $M_{3\pi}\approx 1.3\GeV$ and is
well explained by the so called Deck mechanism \cite{deck} of
diffractive excitation $\pi\to\rho\pi$. The interpretation of the observed peak has been a long
standing controversy, until a phase-shift amplitude analysis (see
references in \cite{pdg}) eventually revealed the presence of the very
weak $a_1$ resonance having a similar mass. 
Moreover, it was found in \cite{belkov} that even the contribution of the $\rho$-$\pi$ cut in the dispersion relation for the diffractive amplitude has a $Q^2$-dependence similar to the $a_1$ pole.
Summing up all
diffractive excitations (excluding large invariant masses
corresponding to the triple-Pomeron term), one concludes that the magnitudes of
single-diffractive and elastic pion-proton cross section are indeed
similar. This helps to resolve the Piketty-Stodolsky puzzle. 

Basing on these observations, in what
follows we employ the simple two-channel model, replacing all heavy 
singularities contributing to the AR, by one effective pole $a$ representing
$a_1$, $\rho-\pi$, etc. We assume that 
\beq 
\sigma_{sd}^{\pi p}(\pi p\to ap)=\sigma^{\pi p}_{el}, 
\label{130} 
\eeq 
and this allows the AR to hold. Notice that applying the AR to
neutrino-production of the effective state $a$, $\nu+p\to l+a+p$, we
should also conclude that \beq \sigma_{tot}^{ap}=\sigma_{tot}^{\pi
p}. \label{135} \eeq
We also assume that the same impact parameter dependences of the elastic $\pi\to\pi$, $a\to a$ and diffractive $\pi\to a$ amplitudes.

At this point we do not pursue a high accuracy of the dispersion
approach, which needs much more model dependent information 
about many singularities contributing to the AR. 
Our objective here is to highlight the importance of absorptive corrections which affect differently the
diagonal and off-diagonal terms in the hadronic current (\ref{100}), which results in an unavoidable 
breakdown of the AR. The proposed simple model, which may be not accurate numerically, provides 
an excellent playground for study of the effects of absorptive corrections keeping physics transparent,
and also allows to estimate the
magnitude of the absorptive corrections.

\section{Diagonal vs off-diagonal diffraction}\label{sect3}

The relation (\ref{130}) between off-diagonal and diagonal
diffractive cross sections cannot be universal and independent of energy and target.
This can be understood within the general
quantum-mechanical interpretation of diffraction
\cite{fp,gw,kl78,mp,kst2,brazil}. A hadron has a composite
structure, and its light-cone wave function consists of  different
hadronic components, the Fock states, which interact with the target differently, what
leads to a modification of their weights. Such a modified wave
packet is not orthogonal any more to other hadrons, what makes possible production
of new hadrons.

It turns out that the off-diagonal diffractive
amplitude can be expressed in terms of diagonal ones. Let us consider two different sets of
states, one consisting of the mass matrix eigenstates, $|h\ra$, and
another one of the states $|\alpha\ra$, which are eigenstates of the
interaction Hamiltonian, i.e. satisfy the condition, $\hat
f_{el}|\alpha\ra = f_\alpha\,|\alpha\ra$, where $\hat f_{el}$ is the
elastic amplitude operator.

Both sets of states are assumed to be complete, so one of them can
be expanded over the full basis of states in the alternative representation,
 \beq
|h\ra = \sum\limits_{\alpha=1}C^h_{\alpha}\,|\alpha\ra.
\label{700}
 \eeq
Due to completeness and orthogonality of each set of these states, the
coefficients $C^h_{\alpha}$ in (\ref{700}) satisfy the relations,
 \beqn
\la h'|h\ra  &=&
\sum\limits_{\alpha=1}(C^{h'}_{\alpha})^*C^h_{\alpha} =
\delta_{hh'}
\nonumber\\
\la \beta|\alpha\ra  &=&
\sum\limits_{h'}(C^{h'}_{\beta})^*C^{h'}_{\alpha} =
\delta_{\alpha\beta}
\label{800}
 \eeqn

The elastic and single diffraction amplitudes can therefore be expressed via
the eigenamplitudes as,
 \beqn
f_{el}^{h\to h} &=& \sum\limits_{\alpha=1}|C^h_{\alpha}|^2\,f_\alpha
\nonumber\\
f_{sd}^{h\to h'} &=&
\sum\limits_{\alpha=1}(C^{h'}_{\alpha})^*C^h_{\alpha}\,f_\alpha
\label{900}
 \eeqn
These relations show that diagonal and off-diagonal diffractive amplitudes are affected by the unitarity (or absorption) corrections quite differently.
For instance, in the black disk limit, which is expected to be reached in the Froissart regime at very high energies
(or in central collisions with a heavy nucleus), all the partial eigenamplitudes reach the
unitarity bound, ${\rm Im}\,f_\alpha=1$. Then, according
to the completeness and orthogonality conditions Eqs.~(\ref{900}), the diffractive amplitudes in the black
disk limit read,
 \beqn
f_{el}^{h\to h}
&\Rightarrow&
\sum\limits_{\alpha=1}|C^h_{\alpha}|^2=1
\nonumber\\
f_{sd}^{h\to h'} &\Rightarrow&
\sum\limits_{\alpha=1}(C^{h'}_{\alpha})^*C^h_{\alpha}
=0
\label{920}
 \eeqn

Off-diagonal diffraction is impossible within a black disc, and may only happen on its periphery,
$b\sim R$. Since in the Froissart regime the interaction radius rises with energy as $R\propto \ln(s)$,
the elastic and diffractive cross sections, which are the amplitudes squared integrated over impact parameter,
acquire different energy dependence,
 \beqn
\sigma_{el}&\propto&
\ln^2(s)
\nonumber\\
\sigma_{sd}&\propto& \ln(s)\ ,
\label{940}
 \eeqn
  i.e. $\sigma_{sd}/\sigma_{el}\propto 1/\ln(s)$, apparently breaking relation (\ref{130}).
Similarly, off-diagonal diffraction on a heavy nucleus is also suppressed,
\beq
\frac{\sigma_{sd}^A}{\sigma_{el}^A} \propto A^{-1/3}.
\label{960}
\eeq

Thus, we conclude that the relation (\ref{130}), which follows from AR, can be strongly broken
either at high energies, or on nuclear targets.

\section{Absorptive corrections to the Adler relation on the proton}\label{sect4}

As was argued above, the absorptive corrections break down validity of Eq.~(\ref{130}).
In order to estimate the magnitude of deviation from the AR on
a proton target, we rely on the simple Regge model proposed in Sect.~\ref{sect2}, with two channels in the axial current, the pion pole
and the effective axial-vector pole $a$ representing the $a_1$ pole  and other singularities producing
the bump at $M\approx 1.3\GeV$ in the invariant mass distribution of the $3\pi$ diffractive excitation of a pion.
As a starting point, we assume that in the single-Pomeron approximation the AR holds, i.e. the diffractive and elastic
amplitudes are equal, i.e. Eq.~(\ref{130}) holds.
Now we introduce absorptive corrections related to the multiple Pomeron exchanges the in initial and
final states, and see how much the relation (\ref{130}) is broken for the output amplitudes.
This can be considered as an estimate for the magnitude of deviation from the AR on a proton.

We rely on the same two-channel model for multi-Pomeron corrections. The unitarized elastic cross section reads,
\beq
\sigma^{\pi p}_{el}=\int d^2b\,
\left[1-e^{-{\rm Im} f^{\pi p}_{el}(b)}\right]^2.
\label{1020}
\eeq
Although this expression looks like the conventional single-channel Glauber model, it remains unchanged within the two-channel model under consideration. For the sake of simplicity we neglect the real part of the amplitude.

Similarly the cross section of the diffractive excitation $\pi p\to a p$, corrected for absorption,
reads,
\beqn
\sigma_{sd}^{\pi p} &=& \int d^2b\,
\left|f_{sd}^{\pi p}(b)\right|^2\,
\left[\frac{e^{-{\rm Im} f^{\pi p}_{el}(b)}-e^{-{\rm Im} f^{a p}_{el}(b)}}
{{\rm Im} f^{a p}_{el}(b)-{\rm Im} f^{\pi p}_{el}(b)}\right]^2
\nonumber\\ &\approx&
\int d^2b\,
\left[{\rm Im} f^{\pi p}_{el}(b)\right]^2\,e^{-2{\rm Im} f^{\pi p}_{el}(b)}.
\label{1040}
\eeqn
At the last step here we used the relation (\ref{135}).

We see that the unitarity corrections to the elastic cross section, Eq.~(\ref{1020}),
and the absorptive corrections to diffraction, Eq.~(\ref{1040}), act in opposite directions:
they enhance the diagonal, but suppress the off-diagonal diffractive processes.

\begin{figure}[htb]
\begin{center}
 \includegraphics[height=6cm]{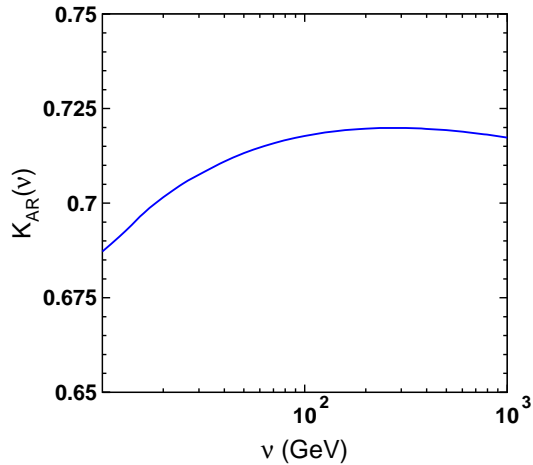}\hspace{10mm}
 \end{center}
\caption{\label{proton-corr} (Color online) The absorptive correction factor for the Adler relation for diffractive neutrino-production of pions on protons.}
 \end{figure}

To estimate the magnitude of the difference between the cross sections (\ref{1020}) and (\ref{1040})
we employ the conventional Gaussian form of the impact parameter dependence for the input single-Pomeron
elastic partial amplitude,
\beq
\Im f_{el}^{\pi p}(b)=\frac{\sigma_{tot}^{\pi p}}{4\pi\,B_{el}^{\pi p}}\,
\exp\left(-\frac{b^2}{2B^{\pi p}_{el}}\right).
\label{980}
\eeq
Then we can evaluate the correction factor $K_{AR}=\sigma_{sd}^{\pi p}/\sigma_{el}^{\pi p}$, which should
be applied to the right-hand side of Eq.~(\ref{40}).
We used (\ref{1020}), (\ref{1040}) and (\ref{980}) with $\sigma_{tot}^{\pi p}=13.6\mb\times s^{0.08}+19.24\mb\times s^{-0.458}$, $B^{\pi p}_{el}=B_0+2\alpha_{\Pom}^\prime\ln s$,
where $s$ is in $\GeV^2$, $B_0=6\GeV^{-2}$, $\alpha_{\Pom}^\prime=0.25\GeV^{-2}$.
The results are depicted in Fig.~\ref{proton-corr}
as function of $\nu$.
We see that the absorptive corrections cause a deviation from the AR  of about $30\%$, which is not a
dramatic effect. This is because the $\pi$-$p$ elastic amplitude is still far from the unitarity bound. However, at
much higher energies (still unreachable in neutrino experiments) the correction factor is expected to
drop significantly.

\section{Coherent neutrino-production of pions on nuclei}

According to the conventional terminology, coherent production on nuclei is a process which leaves the nucleus intact.
Correspondingly, in an incoherent process the nucleus is supposed to break up to fragments.

In what follows we assume the validity of the AR for a nucleon target (unless specified), in order to identify the net
nuclear effects for the AR.
This section is devoted to coherent diffractive pion production. The production amplitudes on different
nucleons interfere, and the interference is enhanced by the condition that the nucleus remains in the ground state.
Such effects of coherence can lead to substantial deviations from the AR and from simplified expectations,
as is demonstrated below.

\subsection{Important time scales}

There are several length scales characterizing the coherence effects in diffractive neutrino scattering on nuclei.
The first length scale is controlled by the longitudinal momentum transfer $q_L^\pi$ in diffractive production
of a pion by an axial current of energy $\nu$ and virtuality $Q^2$,
\beq
l_c^\pi=\frac{1}{q_L^\pi}=
\frac{2\,\nu}{Q^2+m_\pi^2}.
\label{1060}
\eeq
Within this distance the pion production amplitudes $\nu N\to l\pi N$ on different nucleons interfere and shadow
each other.
If the axial current virtuality is small $Q^2\sim m_\pi^2$, the coherence length $l_c^\pi$ is rather long
even at low energies. It reaches the size of a heavy nucleus at energies as low as several hundreds MeV.
Such an early onset of coherence is a peculiar feature of the
axial current.  It would be impossible for the vector current,even for real photoproduction of $\rho$-mesons the onset of shadowing is delayed up to the energies of several GeVs. 
Notice that besides diffractive neutrino-production of pions, the early onset of shadowing also occurs for the total
neutrino-nucleus cross section \cite{k-shad,gransasso,shad-jetp} at low $Q^2$.

Another, much shorter length scale corresponds to diffractive transitions between the axial current and heavy  states,
which are represented by the effective axial-vector meson $a$ in our model,
\beq
l_c^a=\frac{1}{q_L^a}=
\frac{2\,\nu}{Q^2+m_a^2}.
\label{1070}
\eeq
This coherence length controls neutrino diffractive dissociation to heavy hadronic states, and also the
energy dependence of absorptive corrections to the cross section of neutrino-production of pions on nuclei
(see below). At small $Q^2\lsim m_\pi^2$ it is two orders of magnitude shorter than $l_c^\pi$.

\subsection{The amplitude}

The process of coherent neutrino-production of pions on nuclei, $\nu A\to l\pi A$, is possible only
if $l_c^\pi\gsim R_A$, otherwise the nuclear form factor suppresses the cross section  \cite{belkov}.
Even if $l_c^\pi$ is long, the second
length scale $l_c^a$ might be still short. Correspondingly there are few energy regimes where the coherent
length scales vary from very short up to much longer values than the nuclear radius.
Correspondingly, the imaginary part of the partial amplitude of coherent production of a pion contains two terms,
 \beq
 M_{\nu A\to l\pi A}(\nu,Q^2,b)=M_1(\nu,Q^2,b)-M_2(\nu,Q^2,b),
 \label{140}
 \eeq
 where
 \begin{widetext}
 \beqn
 M_1(\nu,Q^2,b)&=&M_{\nu N\to l\pi N}
(\nu,Q^2) \int\limits_{-\infty}^\infty dz\,
e^{iq_L^\pi z}\,
  \rho_A(z,b)\,e^{-{1\over2}\sigma_{tot}^{\pi N}T_A(b,z)};
 \label{160}\\
 M_2(\nu,Q^2,b) &=&
 M_{\nu N\to laN}(\nu,Q^2)\,
 M_{aN\to \pi N}(\nu)
 \int\limits_{-\infty}^\infty dz\,
 e^{i(q_L^\pi-q_L^a) z}\,
  \rho_A(z,b)\,e^{-{1\over2}\sigma_{tot}^{\pi N}T_A(b,z)}
  \nonumber \\&\times&
 \int\limits_{-\infty}^z dz_1 e^{iq_L^a z_1}\,\rho_A(z_1,b)\,
e^{-{1\over2}\sigma_{tot}^{a N}[T_A(b,z_1)-T_A(b,z)]}.
 \label{200}
 \eeqn
\end{widetext}
Here the first term $M_1$ is the amplitude of pion production at the point with longitudinal coordinate
$z$ and impact parameter $b$, integrated over $z$ weighed by the nuclear density $\rho_A$ \cite{belkov}.
The $z$-dependent nuclear thickness function is defined as
\beq
T_A(b,z)=\int\limits_z^\infty dz'\,\rho_A(b,z'),
\label{180}
\eeq
and we denote $T_A(b)\equiv T_A(b,z\to -\infty)$.

 The second term $M_2$ corresponds to diffractive production of the heavy state $a$ preceding the pion production.
This is the first order Gribov inelastic shadowing correction \cite{gribov} to the coherent pion production amplitude.

As far as the amplitude Eq.~(\ref{140}) is known, we can calculate the cross section of coherent pion production,
\beq
\frac{d\sigma(\nu A\to l\pi A)}{dQ^2\,d\nu\,d^2p_T}=
\left|\int \frac{d^2b}{2\pi}\,e^{i\vec p_T\cdot\vec b}M_{\nu A\to l\pi A}(\nu,Q^2,b)\right|^2,
\label{210}
\eeq
where $\vec p_T$ is the transverse momentum transfer to the target, and we neglect the real part of the ampltude.

The cross section on a nucleon target according to (\ref{980}) has the form,
\beq
\frac{d\sigma(\nu N\to l\pi N)}{dQ^2\,d\nu\,d^2p_T}=
\frac{e^{-B_{el}^{\pi p}p_T^2}}{(2\pi)^2}\,
\left|M_{\nu N\to l\pi N}(\nu,Q^2)\right|^2
\label{215}
\eeq

\subsection{Characteristic regimes in the energy dependence}

In the general expression Eq.~(\ref{140})-(\ref{200}) one can identify several characteristic regimes,
which are controlled by the interplaying coherence scales $l_c^\pi$ and $l_c^a$.

 \subsubsection{\boldmath$l_c^\pi\lesssim R_A$}\label{sub1}

 In this regime the AR on a nucleus trivially breaks down, as one can see from Eq.~(\ref{160}). The cross section
 is falling with decreasing $l_c^\pi$ and vanishes at $l_c^\pi\ll R_A$, as it was calculated in \cite{belkov}.
 The reason is obvious: the AR relation is supposed to hold at the pion pole at $Q^2=-m_\pi^2$,
 and the extrapolation to $Q^2=0$ leads to a strong variation of the amplitude, if the longitudinal momentum transfer $l_c^\pi$ is comparable with $R_A$, or shorter.
 
Notice that one should be cautious applying Eq.~(\ref{160}) at low energies where the neglected contribution of  s-channel resonances and/or reggeons is important \cite{khlopov1,khlopov2}. Also the neglected real part of the amplitude becomes large.
Therefore, our calculations for this energy regime of  $l_c^\pi\lesssim R_A$ only present an estimate of  the effects 
related to the nuclear formfactor \cite{belkov}.

\subsubsection{\boldmath$l_c^\pi\gg R_A$, $l_c^a\ll R_A$}\label{sub2}

Eqs.~(\ref{160})-(\ref{200}) are significantly simplified in this regime of very long $l_c^\pi$ and very short lifetime
$l_c^a$ of the $\pi\to a$ fluctuations compared to the nuclear radius $R_A$.
In this case the amplitude $M_2(\nu,Q^2,b)$, Eq.~(\ref{200}) is strongly suppressed by
the oscillating exponential and can be neglected.  At the same time,  $l_c^\pi\gg R_A$, therefore
the non-vanishing term in (\ref{140}), the amplitude $M_1(\nu,Q^2,b)$, can be also simplified by integrating over
$z$ in Eq.~(\ref{160}) analytically,
\beq
M_1(\nu,Q^2,b)=
 \frac{2M_{\nu N\to l\pi N}(\nu,Q^2)}{\sigma_{tot}^{\pi N}}\,
 \left[1-e^{-{1\over2}\sigma_{tot}^{\pi N}T_A(b)}\right]
 \label{220}
 \eeq

 At this point we concentrate on nuclear effects leading to breakdown of the AR, therefore hereafter we
 assume that the amplitude of pion neutrino-production on a nucleon satisfies the AR Eq.~(\ref{40}),
 \beq
 M_{\nu N\to l\pi N}(\nu,Q^2=0)=
 \xi(E,\nu)\,{1\over2}\sigma^{\pi N}_{tot}(\nu),
 \label{240}
 \eeq
 where we employ the optical theorem neglecting the real part of the diffractive amplitude.
 The amplitudes are normalized as $d\sigma/dQ^2d\nu=\left|M_{\nu N\to l\pi N}(\nu,Q^2)\right|^2$.
 Applying (\ref{180}) to (\ref{160}) we get the relation,
 \beq
 M_1(\nu,Q^2=0,b)=
 \xi(E,\nu)\,\left[1-e^{-{1\over2}\sigma_{tot}^{\pi N}T_A(b)}\right],
 \label{260}
 \eeq
which is exactly the AR for the partial amplitude of neutrino-production of pions.
We conclude that if the AR is correct for a pion production on a nucleon, it should be also correct
for a nuclear target, provided that $l_c^\pi\gg R_A$, but $l_c^a\ll R_A$.

  \subsubsection{\boldmath$l_c^a\gg R_A$}\label{sub3}

In this regime all the phase shifts in Eq.~(\ref{140}) can be neglected and the integration over
$z$ and $z_1$ can be performed analytically,
 \begin{widetext}
  \beqn
 M_2(\nu,Q^2,b)\Bigr|_{l_c^a\gg1} &=&
M_{\nu N\to aN}(\nu,Q^2)\,
 M_{aN\to \pi N}(\nu)\,
 \frac{4}{\sigma_{tot}^{a N}}
\Biggl\{\frac{1}{\sigma_{tot}^{\pi N}}\,
 \left[1-e^{-{1\over2}\sigma_{tot}^{\pi N}T_A(b)}\right] \,
-\,  \frac{e^{-{1\over2}\sigma_{tot}^{a N}T_A(b)} -
  e^{-{1\over2}\sigma_{tot}^{\pi N}T_A(b)}}
  {\sigma_{tot}^{\pi N}-\sigma_{tot}^{a N}} \Biggr\}
  \nonumber\\ &=&
M_{\nu N\to aN}(\nu,Q^2)\,
 M_{aN\to \pi N}(\nu)\,
 \frac{4}{\left(\sigma_{tot}^{a N}\right)^2}
\Biggl\{1-\left[1+{1\over2}\sigma_{tot}^{\pi N}T_A(b)\right]
e^{-{1\over2}\sigma_{tot}^{\pi N}T_A(b)}
 \Biggr\}
 \label{280}
\eeqn
 \end{widetext}

The first term in (\ref{140}), the amplitude $M_1$, was already calculated in this limit ($l_c^a\gg R_A$)
above in (\ref{220}).
In both expressions the neutrino-production amplitudes $M_{\nu N\to hN}(\nu,Q^2)$ are related to the hadronic ones,
$a+N\to h+N$, by the AR . Besides, based on the assumed dominance in the axial current of the effective pole $a$,
we can extrapolate these relations to nonzero $Q^2$ with the pole propagator $(Q^2+m_a^2)^{-1}$. Thus, we get
new relations,\\

\beq
M_{\nu N\to laN}(\nu,Q^2)=\frac{\xi(\nu)\,m_a^2}{Q^2+m_a^2}\,M_{aN\to a N};
\label{290}
\eeq
\beq
M_{\nu N\to l\pi N}(\nu,Q^2)=\frac{\xi(\nu)\,m_a^2}{Q^2+m_a^2}\,M_{aN\to \pi N},
\label{300}
\eeq
 We rely on these relations in what follows.
 
 Eventually, summing the amplitudes Eqs.~(\ref{220}) and (\ref{280}) we arrive at
 \beq
 \frac{M_{\nu A\to l\pi A}(\nu,Q^2,b)}
 {M_{\nu N\to l\pi N}(\nu,Q^2)}\Biggr|_{l_c^a\gg1} =
 T_A(b)\,e^{-{1\over2}\sigma_{tot}^{\pi N}T_A(b)}.
\label{320}
\eeq

 One can observe the striking difference in the $A$ dependence of the cross sections corresponding to
 the regimes of short
 and long coherence length $l_c^a$, equations (\ref{220}) and (\ref{320}) respectively.
 The AR holds in the former case and nuclear cross section behaves as $A^{2/3}$ (for very heavy nuclei).
However, at latter case the cross section is $\propto A^{1/3}$, so the AR breaks down. 
 Notice that in this regime of $l_c^a\gg R_A$ the calculations \cite{belkov} based on AR are not correct. Also the results of \cite{rs} are not correct at any energy, since rely on a wrong model for the pion-nucleus cross section (see discussion in \cite{belkov}).

 \subsection{Numerical results}\label{sect:numerical}

\begin{figure}[htb]
\begin{center}
 \includegraphics[height=7cm]{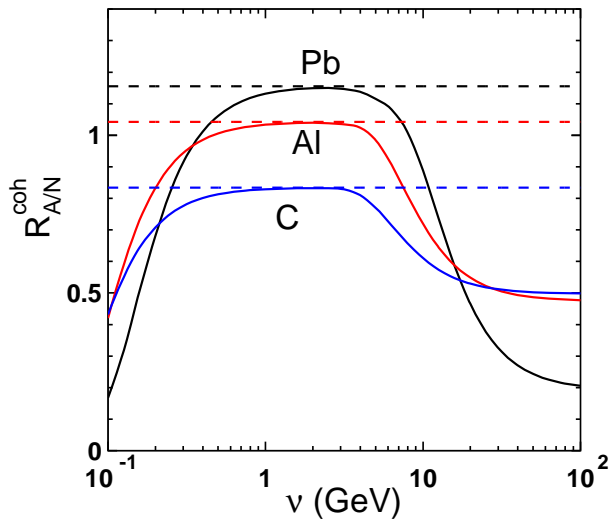}\hspace{10mm}
 \end{center}
\caption{\label{fig:coh-Adep} (Color online) {\it Solid curves:} nuclear ratio $R^{coh}_{A/N}(\nu,Q^2)$ of $p_T$-integrated cross sections of coherent neutrino-production of pions, $\nu+A\to l+\pi+A$,
calculated with Eq.~(\ref{340}) at $Q^2=0$. {\it Dashed lines:} the results of the
Adler relation applied to nuclear targets, lead, aluminium and carbon, from top to bottom.}
 \end{figure}

 Such a nontrivial behavior of nuclear effects as function of energy is confirmed by the results of numerical
 calculations of the $p_T$-integrated cross sections with Eqs.~(\ref{140})-(\ref{200}). The ratio
\beqn
 && R^{coh}_{A/N}(\nu,Q^2) = \frac{d\sigma(\nu A\to l\pi A)/dQ^2\,d\nu}
 {A\,d\sigma(\nu N\to l\pi N)/dQ^2\,d\nu}
 \nonumber\\ &=&
 \frac{4\pi B^{\pi N}_{el}}{A}\,
\int d^2b \left|M_{\nu A\to l\pi A}(\nu,Q^2,b)\right|^2
 \label{340}
\eeqn
 is plotted in Fig.~\ref{fig:coh-Adep} for lead, aluminium and carbon targets as function of $\nu$ at $Q^2=0$.
  For the sake of simplicity the calculations were performed with a constant cross section $\sigma_{tot}^{\pi N}=25\mb$.

 We see that at low energies $\nu\lesssim 1\GeV$ the nuclear ratio $R^{coh}_{A/N}$, plotted by solid curves,
 rises with $\nu$ and saturates at the level corresponding to the AR applied to nuclei, depicted by dashed
 horizontal lines. Because the survival probability of a pion propagating a long path length in the nuclear
 medium is low, the pion production points
are pushed to the back surface of the nucleus (compare with Eq.~(\ref{220})). Therefore, the cross section
depends on nuclear atomic number as $\sim A^{2/3}$ and coincides with the prediction of AR in the saturated
regime of $q_c^\pi\ll1/R_A$.

 The observed strong deviation from the AR prediction at very low energies is a simple consequence of the
 suppression of the coherent cross section caused by the nuclear formfactor due to finiteness of the  momentum
 transfer, $q_c^\pi\sim 1/R_A$.
 
  The energy dependence of nuclear ratio forms a plateau from several hundreds MeV up to several GeV in the regime described in Sect.~\ref{sub2}.
 It also well agrees with the prediction of the AR shown by dashed lines in Fig.~\ref{fig:coh-Adep}.

 At energies $\nu\gsim10\GeV$ the nuclear cross section considerably drops and saturates at a new level exposing a significant deviation from the expectations based on the AR, depicted by dashed curves. This happens due to the transition to the new regime of full coherence explained in Sect.~\ref{sub3}.

 It worth commenting that the height of the plateaus for different nuclei shows that the $A$-dependence of the cross section in this regime is slightly steeper than linear.
This is different from the simple expectation of  $R^{coh}_{A/N}\propto A^{-1/3}$ corresponding to  the black disc limit. This happens because the cross section $\sigma^{\pi N}_{tot}$ is rather small and the pion-nucleus
 partial amplitude is still far from the unitarity bound. So the pion-nucleus elastic cross section
 is quite smaller than the $\pi R_A^2$. This is why it rises as $A^\alpha$
 with $\alpha>1$. 

Thus, the cross section of diffractive coherent neutrino-production of pions on nuclei exposes a peculiar 
energy dependence. It starts from zero at very small energies, then rises and saturates  at a large magnitude, and eventually drops down to a value $\propto A^{1/3}$ at higher energies. The AR relation
is severely broken at the regimes of short $l_c^\pi\ll R_A$ and long $l_c^a\gg R_A$, but is rather accurate within
the intermediate regime.

\begin{figure}[htb]
\begin{center}
 \includegraphics[height=7cm]{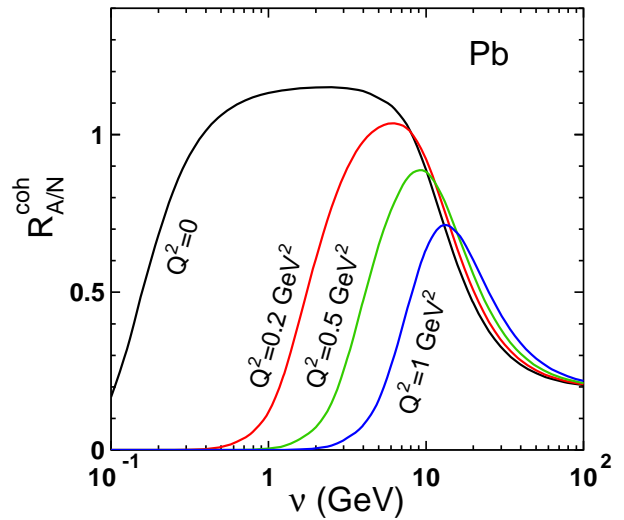}\hspace{10mm}
 \end{center}
\caption{\label{fig:coh-Qdep} (Color online) The same as in Fig.~\ref{fig:coh-Adep} for lead  at $Q^2=0,\ 0.2,\ 0.5$ and $1\GeV^2$.}
 \end{figure}

 The specific energy dependence of nuclear effects presented in Fig.~\ref{fig:coh-Adep} at $Q^2=0$ drastically
 changes with rising $Q^2$. Indeed, the plateau in the energy dependence, which spans across the wide energy range,
 is related to the significant difference between the length scales Eqs.~(\ref{1060})-(\ref{1070}), $l_c^\pi\gg l_c^a$.
 This holds, however, only for tiny values of $Q^2\lesssim m_\pi^2$. With rising $Q^2$ both scales contract  down
 to the same order of magnitude, and the plateau in the energy dependence of $R_{A/N}^{coh}$ shrinks and becomes a peak.
 This is illustrated in Fig.~\ref{fig:coh-Qdep} for neutrino-production of pions on lead for few values of
 $Q^2=0,\ 0.2,\ 0.5$ and $1\GeV^2$.

We do not extend our predictions to larger values of $Q^2$ for several reasons. First of all, at large $Q^2$ the
effects of color transparency make the nuclear medium more transparent than we evaluated. These effects cannot be
reproduced within the employed two-channel model. In hadronic
representation color transparency results from superposition of many singularities in the dispersion relation with masses up to $M^2\sim Q^2$ \cite{zkl,jk}.
Besides,
one should take care of the correct (negative) signs of the off-diagonal diffractive amplitudes, and provide a fine
tuning
between different amplitudes, which must essentially cancel each other at high $Q^2$, in order to end up with color
transparency. This is a difficult task, which can be solved much more effectively within the dipole representation
\cite{zkl}.

Another reason for not extending our calculation to larger values of $Q^2$ is the missed contributions of the
transverse component of the axial current and of the vector current. Both vanish at $Q^2\to 0$,
but should be added and have growing importance with rising $Q^2$.

 We presented numerical results for nuclear effects only for $p_T$-integrated cross sections, since their
 $p_T$-dependence is rather simple and well known.
 The $p_T$-distribution of coherent pion production forms a narrow peak at small $p_T$, with a slope of the order
 of ${1\over3}R_A^2$, caused by the nuclear form factor. More accurately, the $p_T$-dependence of the cross section
 is given by Eq.~(\ref{210}).
 The large $p_T$-slope of the cross section is the signature of the coherent process, which is usually used to
 disentangle it from the incoherent background having a much smaller slope, similar to production on a free nucleon.

 \section{Incoherent pion production}\label{sect:incoh}

 As a result of momentum transfer in diffractive neutrino-production on a bound nucleon, the nucleus can be
 excited or break-up to fragments, $\nu+A\to l+\pi+A^*$. Although pions diffractively produced at different impact parameters
 do not interfere in this process, characterized by rather large transverse momentum transfer, the amplitudes on
 bound nucleons with the same impact parameter do interfere.
 Evaluation of the cross section is more involved that in the case of coherent production, but can be simplified
 by summing up all nuclear final states and employing completeness.
 We perform calculations within the two-channel model for the axial current introduced earlier.
 The results are presented in the form of nuclear ratio defined similar to the case of coherent production, Eq.~(\ref{340}), 
  \beq
  R^{inc}_{A/N}(\nu,Q^2) = \frac{d\sigma(\nu A\to l\pi A^*)/dQ^2\,d\nu}
 {A\,d\sigma(\nu N\to l\pi N)/dQ^2\,d\nu}.
\label{360}
\eeq

\subsection{Effects of coherence for incoherent production}

Like in the case of coherent production, one can identify several contributions in the nuclear factor
$R^{inc}_{A/N}$, characterized with different mechanisms \cite{hkn-glaub}.
 \beq
 R^{inc}_{A/N}=R^{inc}_1+R^{inc}_2-R^{inc}_3.
 \label{370}
 \eeq
 The three terms in the right-hand side of this equation correspond to the following mechanisms of incoherent pion production.
 
 I) The incoming neutrino does not interact in the nucleus up to the point with coordinates $(b,z)$, where it
diffractively produces the pion, $\nu+N\to l+\pi+N$, which survives propagating through the nucleus.

The corresponding amplitude squared, summed over the final state of the nucleus, and integrated over coordinates
of the bound nucleon has the form,
  \beqn
R^{inc}_1&=& {1\over A}\
 \int d^2b \int\limits_{-\infty}^{\infty}
 dz\
\rho_A(b,z)\,e^{-\sigma^{\pi N}_{in} T_A(b,z)}
\nonumber\\ &=&
\frac{1}{A\sigma^{\pi N}_{in}}
\int d^2b\,\left[1-e^{-\sigma^{\pi N}_{in} T_A(b)}\right]
\label{380}
\eeqn

 II) Prior the pion production  the neutrino interacts with another bound nucleon at the point ($b,z_1$),
 and produces diffractively an $a$-meson, $\nu+N\to l+a+N$, which is the effective state representing different products of diffractive excitation of a pion, as it was introduced in Sect.~\ref{sect2}.
Then the $a$-meson propagates further and produces diffractively a pion, $a+N\to\pi+N$ ($z>z_1$).
The corresponding term in the nuclear factor derived in \cite{hkn-glaub} has the form,
\begin{widetext}
 \beqn
R^{inc}_2 &=&
\frac{\sigma^{\pi N}_{tot}}
 {2A\,\sigma^{\pi N}_{el}}\
 (\sigma^{\pi N}_{in} -
\sigma^{\pi N}_{el})\,\int d^2b
 \int\limits_{-\infty}^{\infty} dz_1\
 \rho_A(b,z_1)\
\int\limits_{z_1}^{\infty} dz_2\
 \rho_A(b,z_2)\,
 \cos\left[q_c^\pi(z_2-z_1)\right]\
 \nonumber\\
 & \times&
 \exp\left[- {1\over 2}(\sigma^{\pi N}_{in} -
\sigma^{\pi N}_{el})\ T_A(b,z_2) -
 {1\over 2}\sigma^{\pi N}_{tot} T_A(b,z_1)
\right]
\label{400}
\eeqn
\end{widetext}
Here we fixed $\sigma^{a N}_{tot}=\sigma^{\pi N}_{tot}$, as  follows from the AR
in the employed two-channel model; $z_1$ and $z_2$ are (\ref{400}) the longitudinal coordinates  of diffractive
neutrino-production of the intermediate $a$-meson in the two interfering amplitudes. The final pion is produced
diffractively, $a+N\to\pi+N$, but incoherently, i.e. on the same nucleon, with coordinates $(b,z)$ in both amplitudes.

III. In the first two terms of (\ref{370}) we summed up all final state of the nucleus including the ground state.
The latter corresponds to coherent pion production evaluated in the previous section, and should be subtracted.
Thus,
 \beq
 R^{inc}_3 =  \frac{\left(\sigma^{\pi N}_{tot}\right)^2}{4A\,\sigma^{\pi N}_{el}}
\int d^2b
\left| \int\limits_{-\infty}^{\infty}
 dz\,\rho_A(b,z)\ e^{iq_c^\pi z}
 e^{-{1\over
2}\sigma^{\pi N}_{tot}\ T_A(b,z)}
 \right|^2
\label{420}
\eeq

Like in the case of coherent production, one can identify three regimes of energy dependence of the incoherent cross section. 

\subsubsection{\boldmath$l_c^\pi\lsim R_A,\ \ l_c^a\ll R_A$}

 In the low energy limit of $q^\pi_c\gg R_A$ only the first term in (\ref{370}) survives and
 $R^{inc}_{A/N}\Bigr|_{q^\pi_c\gg R_A}=R^{inc}_1$ given by Eq.~(\ref{380}).

 At higher energies, when $q^\pi_c\to 0$ all integrations on longitudinal coordinates in
 (\ref{380})-(\ref{420}) can be performed analytically,
 \beq
 R^{inc}_{A/N}\Bigr|_{q_c^\pi \to0}=\int d^2b\,T_A(b) e^{-\sigma^{\pi N}_{in} T_A(b)}.
 \label{440}
 \eeq
 This shows a considerable drop of the nuclear ratio from the low energy limit given by Eq.~(\ref{380}) toward the high energy limit.
 The interpolation between the two regimes is performed by the full expression Eqs.~(\ref{340}).
 
\begin{figure}[htb]
\begin{center}
 \includegraphics[height=7cm]{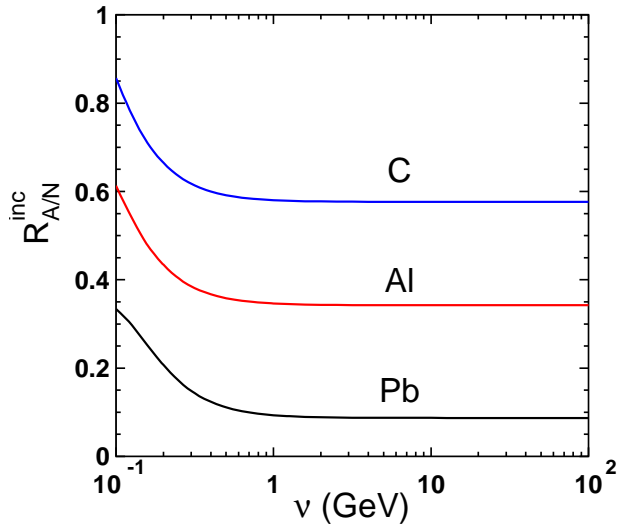}\hspace{10mm}
 \end{center}
\caption{\label{fig:inc-Adep} (Color online) The same as in Fig.~\ref{fig:coh-Adep} for incoherent pion production $\nu+A\to l+\pi+A^*$.}
 \end{figure}

 The numerical results at $Q^2=0$ for several nuclei depicted in Fig.~\ref{fig:inc-Adep} indeed demonstrate a considerable drop with energy of the nuclear ratio.
 Notice that a similar behavior predicted in \cite{hkn-glaub} for electroproduction of vector mesons,
was nicely confirmed later the by HERMES experiment \cite{hermes} (see also \cite{knst}).

\begin{figure}[b]
\begin{center}
 \includegraphics[height=7cm]{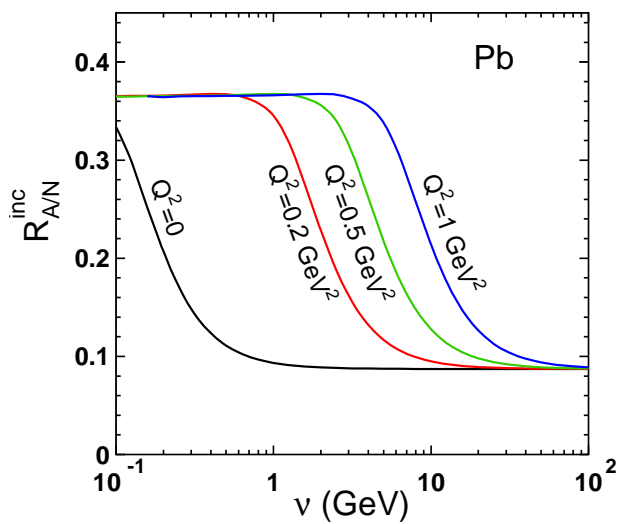}\hspace{10mm}
 \end{center}
\caption{\label{fig:inc-Qdep} (Color online) The same as in Fig.~\ref{fig:inc-Qdep} for lead  at $Q^2=0,\ 0.2,\ 0.5$ and $1\GeV^2$.}
 \end{figure}

At large values of $Q^2$ the regime of short $l_c^\pi$ propagates to higher energies, as is demonstrated
in Fig.~\ref{fig:inc-Qdep}.

So far we assumed that $l_c^\pi$ may be short or long, but the second length scale $l_c^a$ is always short.
In this case, similar to the coherent process in this regime, the AR is valid. Indeed, Eq.~(\ref{440}) is equivalent
to the Glauber formula for nuclear ratio in quasi-elastic pion scattering on a nucleus, i.e. is exactly what
follows from the AR.

\subsubsection{\boldmath$l_c^a\gg R_A$}

At higher energies $l_c^a$ also becomes long, what has led to breakdown of the AR in the coherent process
(see the previous section and Fig.~\ref{fig:coh-Adep}).
What happens in this case with an incoherent pion production ? In the asymptotic regime of $l_c^a\gg R_A$ the
answer is easy,
 \beq
  R^{inc}_{A/N}\Bigr|_{q_c^a\gg R_A}=
  \int d^2b\,\frac{e^{-\sigma^{a N}_{in} T_A(b)}-e^{-\sigma^{\pi N}_{in} T_A(b)}}
  {\sigma_{in}^{\pi N}-\sigma_{in}^{aN}}.
 \label{460}
 \eeq
We have shown above, Eq.~(\ref{135}),  that in the two-channel model under consideration the AR leads to the
equality $\sigma^{a N}_{in}=\sigma^{\pi N}_{in}$. In this case Eq.~(\ref{460}) is equivalent to (\ref{440}).
Thus, we arrived at a remarkable conclusion that in the case of incoherent neutrino-production of pions on
nuclear targets the AR is always correct.

\section{Summary}

At high energies neutrinos expose hadronic properties similar to photons, since they also interact with a
target via hadronic fluctuations. Although it is tempting to interpret the AR as pion dominance, the pion
pole is excluded due to conservation of the leptonic current (for neutral current, otherwise is suppressed
by the lepton mass).
In fact, the AR imposes a mysterious relation between the pion interaction with the target and the contribution
of heavy axial states to the neutrino interaction. The former corresponds to elastic pion scattering in the
process of diffractive neutrino-production of pions, while the latter is related to off-diagonal diffraction
of a pion, excluding elastic scattering.
It is known that these two processes are subject to absorptive corrections which affect them quite differently,
namely, they enhance diagonal diffraction (elastic scattering), but suppress inelastic diffraction. Therefore,
the AR cannot be universal, target independent.

We checked the role of absorptive corrections for diffractive neutrino-production of pions on protons and
nuclei. Assuming that the AR holds on a proton target without absorptive corrections, we estimated the magnitude
of deviation from AR at about 30\% (see Fig.~\ref{proton-corr}).

Much stronger effects were found on heavy nuclei. In coherent production of pions, $\nu+A\to l+\pi+A$,
the AR holds with a good accuracy at energies $\nu\approx 1-10\GeV$.  However, it is severely broken at lower and
higher energies (see Fig.~\ref{fig:coh-Adep}). Our numerical results at low energies in the regime of $l_c^\pi\lesssim R_A$ are rather schematic, since do not include the contribution of resonances and large real part of the diffractive amplitudes.   

For incoherent pion production, $\nu+A\to l+\pi+A^*$, when the nucleus decays into fragments, we found a considerable variation of
nuclear effects with energy (see Fig.~\ref{fig:inc-Adep}), similar to photoproduction of vector mesons. Remarkably,
however, no deviations from the AR were detected, and it holds at all energies.

While the employed two-channel model may be numerically not very accurate, it allows to simplify the calculation of the absorptive corrections and estimate the magnitude of deviations from the AR. Besides, explicit involvement of
heavier singularities in the dispersion relation would lead to appearance of many unknown parameters.
An alternative description, which allows to include all of them would be the light-cone color dipole representation 
\cite{zkl}. The corresponding results will be published elsewhere \cite{dipole1}.

\begin{acknowledgments}

We are thankful to Daniel Ega$\tilde{\rm n}$a, Genya Levin and Arkady Vainshtein for useful  discussions. This work was supported in part
by Fondecyt (Chile) grants 1090236, 1090291, 1100287 and 1090073, and by Conicyt-DFG grant No. 084-2009.

\end{acknowledgments}

\end{document}